\documentstyle[prl,aps,epsfig,multicol]{revtex}
\newcommand{\bm}{\bibitem}

\def\be {\begin{equation}}
\def\ee {\end{equation}}
\def\bea {\begin{eqnarray}}
\def\eea {\end{eqnarray}}
\def\nn {\nonumber}

\newcommand {\p} {{\Theta}(E,E^\prime)dE^\prime}
\newcommand {\th}{{\Theta}}
\newcommand {\wpl}{{\omega}}
\begin{document}
\title{Stopping power of hot QCD plasma }
\author{Abhee K. Dutt-Mazumder$^a$, Jan-e Alam$^b$, 
Pradip Roy$^a$, Bikash Sinha $^{a,b}$}
\address{a) Saha Institute of Nuclear Physics, 1/AF Bidhannagar, Kolkata, India}
\address{b) Variable Energy Cyclotron Centre, 1/AF Bidhannagar, Kolkata, India}
\maketitle

\vspace{0.5cm}

\begin{abstract}
The partonic energy loss has been calculated taking both the hard
and soft contributions for all the 
$2 \rightarrow 2$ processes, revealing the importance
of the individual channels. 
Cancellation of the 
intermediate separation scale has been exhibited.  
Subtleties related to the identical final state partons have properly
been taken into account.
The estimated collisional loss is compared with its radiative counter part.
We show that there exists a critical energy ($E_c$) below which
the collisional loss is more than its radiative counterpart. 
In addition, we present closed form formulas for both the collision 
probabilities and the stopping power ($dE/dx$). 
\\[0.1 cm]
{ PACS numbers: 11.15.Bt, 04.25.Nx, 11.10.Wx, 12.38.Mh }
\end{abstract}
\vspace{0.2cm}

\begin{multicols}{2}
\section{Introduction}
The partonic energy loss in a QCD plasma has received 
significant attention in recent years.  Experimentally, the partonic energy 
loss can be probed by measuring the high $p_T$ hadrons emanating from 
ultra-relativistic heavy ion collisions. This idea was first
proposed by Bjorken \cite{bjorken} where `ionization loss' of 
the quarks and gluons in a QCD plasma was estimated. In fact, the `stopping 
power' ($dE/dx$) of the plasma is proportional 
to $\sqrt{\epsilon}$, where, $\epsilon$ 
is the energy density of the partonic medium. Therefore, by measuring 
various high $p_T$ observables one can probe the initial parton density
\cite{bjorken}.

Hard partons, injected into hot QCD medium, can dissipate energy in two
ways, {\em viz.}, by two body collisions or {\em via} the bremsstrahlung 
emission of gluons, commonly referred to as collisional and radiative loss 
respectively.  For electromagnetic processes, it is well known that at large 
energies, radiative losses are much higher than the collisional loss. 
In fact, there is a  critical 
energy $E_c$, at which, both the processes contribute 
equally~\cite{rossi1,rossi2,leo}. 
In QCD plasma, however, the situation is more involved because
of the non-abelian nature of the interaction. 
We shall show that for low energy partons $2 \rightarrow 2$ processes
are equally important. This is particularly so for heavy quarks~\cite{mustafa}. 
In QCD plasma, to our knowledge, such estimation of $E_c$ is not known yet.
We address this issue in the present work. 
Obviously this requires 
complete treatment of both $2 \rightarrow 2$ and $2 \rightarrow
3$ (or higher order) processes. While significant progress has been made
over the past decade to estimate bremsstrahlung induced partonic energy loss
\cite{gyulassy00,gyulassy00prl,gyulassy95,gyulassy94,baier95,baier96,baier97a,baier97b,baier98}, 
collisional loss, as we uncover, begs further attention. 
The energy loss of partons in a QCD plasma 
due to the dissociation of the possible binary bound states in a strongly
coupled QGP has recently been considered by Shuryak 
{\it et al.}~\cite{shuryak}. They have shown that the partonic energy
loss due to  this process  is important 
in a narrow interval of plasma temperatures. 
It may be noted that unlike radiative loss 
in which mostly relativistic gluons are produced, collisional/ionization loss,
on the contrary, 
injects energy-momentum and entropy into the plasma.
However, in this work we 
discuss only parton parton scattering, not the dissociation of the 
binary bound state. 

The partonic energy loss in a QCD plasma was first estimated 
by  Bjorken \cite{bjorken}. Considering two body scattering of the parton
off thermal quarks and gluons, 
the following expression for the energy loss is
obtained. 
\bea
\frac{dE}{dx}=\frac{8\pi}{3}(1+\frac{n_f}{6})\,\alpha_s^2 T^2 
\ln(\frac{q_{\mathrm max}}{q_{\mathrm min}})
\label{bjorken}
\eea
In the above
equation, $n_f/6$ term is for quark sector while the other one is responsible
for the gluonic loss.
Bjorken retains only the infrared divergent part of the integral. 
Here $q_{\mathrm min}$ and $q_{\mathrm max}$ denote maximum and minimum
momentum transfer. Evidently, as indicated in ref.\cite{bjorken} itself, that
this expression breaks down in the infrared region due to plasma effects.
The presence of $q_{max}$ and $q_{min}$ should be noted here, for which
one takes reasonable values from physical argument. In principle,
however, should emerge from the theory itself in a natural way.

Physically energy losses for partons traversing plasma can be divided into two 
parts. One due to close collisions involving hard momentum transfer,
to be treated microscopically in terms of individual scatterings,
and the other for the distant collisions involving soft 
momentum transfer. Evidently, for the latter, the de Broglie wave
length of the exchanged particle becomes comparable with the inter particle
distance ($\sim T^{-1}$), which renders the concept of individual
scattering meaningless and necessitates the inclusion of the plasma
effects.

In the long wave length regime the problem of energy loss can be treated 
macroscopically in terms of classical chromoelectric field. This coherent
regime of partonic energy loss was first calculated by Thoma and Gyulassy
\cite{thoma91} for the heavy quark. They actually express relative energy loss 
in terms of the chromodielectric field tensor by combining hard thermal loop
corrected gluon propagator with the techniques of classical plasma physics
as: 
\bea
&&\frac{dE}{dx}=\frac{C_F\alpha_s}{2\pi^2 v}
\int\frac{\omega d^3k}{k^2}\nn\\
&&\left[
Im\epsilon_l^{-1}
+(v^2k^2-\omega^2)Im(\omega^2\epsilon_t-k^2)^{-1}
\right ] 
\label{gt}
\eea
In this formalism, the ambiguity related with the lower cut off is naturally 
removed by the inclusion of the polarization
effects. However, this description also breaks down at some ultraviolet 
scale \cite{mrow91,braaten91} and should therefore be cut off
at some hard momentum scale.

Evidently, Eqs.~\ref{bjorken} and ~\ref{gt} capture physics relevant
for two different kinematic regime and therefore complimentary to each other.
Any complete calculation should include both the infrared and ultraviolet 
part as discussed at length by Mrowczynski~\cite{mrow91} and 
Braaten {\it et al.}\cite{braaten91} 
to evaluate the total energy loss.
In ref.\cite{mrow91} for the first time a complete  calculation for the 
partonic
energy loss was presented. Here the hard part
was treated along the line of ref.\cite{bjorken} and the soft part was
calculated classically. However, the ambiguity related to the separation of
scale remained till the full field theoretic techniques was developed
by Braaten and Pisarski~\cite{pisarski} 
 which was subsequently  used to calculate the energy 
loss of heavy fermions by Braaten and Yuan~\cite{yuan91}. 
In their formalism it was shown how the infrared
cut off naturally arise from the hard thermal loop resummation scheme and
at the same time the intermediate scale dependence is removed from the
theory where the separation of scale appears in the argument of the logarithms
and cancels automatically when contributions of these two regimes are 
added together~\cite{braaten91,thomaplb91}. 

In retrospect of these developments we revisit the problem of light
quark and gluon energy loss in a QCD plasma. 
The treatment of light quarks (or gluons) is different from that of
heavy quarks in many different ways. Primarily this is related to the
presence of the light quarks or gluons in the thermal bath. This would
modify the thermal phase space. In addition there will be back reactions
which should be taken into account. Moreover, unlike heavy quark, their
light partners can even be annihilated with the thermal constituents.
Most importantly, light quarks or gluons involve subtleties 
related to the processes having identical final state species 
which was not properly taken into account \cite{bjorken}.

The motivation of the present
work is to unravel the relative contribution of individual processes. 
Therefore, we take all possible channels including elastic and 
inelastic scatterings.  Consequently, estimated $(dE/dx)_{\mathrm{coll}}$, 
is higher than reported before. In fact, the two body compton like 
scattering, proves to be quite efficient in transferring energy 
into the plasma.  In addition, we also evaluate explicitly the 
collisional loss of gluon energy. 
For this, $gg \rightarrow gg$ and $q g \rightarrow q g$ are found to be 
most important. 

It might be mentioned that calculations reported in 
\cite{bjorken} (see also~\cite{baier00}) were restricted only to 
the $t$ channel processes,
thereby, excluding the interference and exchange terms, which contribute
significantly. They are particularly important for processes having
$u$ channel divergences.

The plan of the paper is as follows. In section II, we develop the formalism.
which will be used afterwards. In section III contributions from various 
channels on the collisional energy loss of partons have been presented.
These results are then compared with radiative loss. 
In section IV, we explicitly treat both the soft and hard momentum transfer
regime and show how infrared divergence, originally present in the previous
sections are automatically removed by the plasma effects. In addition we
demonstrate that the final expression is free from the momentum cut off
introduced to regulate the divergences in two (soft and hard) kinematic regimes.
Section V is devoted to summary and discussion. Various sum rules used in
section IV are collected in the appendix.

\section{Formalism}

While the heavy quark energy loss is very similar to the muon energy loss
in a plasma of electrons and positrons, light quark energy loss
is analogous to the electron energy loss in a QED plasma.
Therefore to calculate the relevant $dE/dx$ arising out of
two body scatterings, we introduce a formalism along the line similar to what 
is employed to study cosmic ray showers \cite{rossi1,rossi2}. 
Accordingly, we define a differential collision probability, $\p dx$ which
represents the probability of a parton with energy $E$ 
to transfer an amount of energy between $E^\prime$ and $E^\prime + dE^\prime$ 
to a plasma constituent in traversing a thickness $dx$. Energy loss can be 
obtained by convoluting $\p$ with the energy transfer $(E^\prime)$ for each 
processes which generically is given by,  

\bea
\frac{dE}{dx}=
\int_{E_{\mathrm{min}}}^{E_{\mathrm{max}}} E^\prime \p. 
\label{dedx}
\eea

In the above equation, $E_{\mathrm max}$ is the maximum energy transfer, while
$E_{\mathrm min}$ is a cut off used to regulate infrared divergence 
related to the usual small angle limit. However, in case of energy 
loss calculation $\int \frac{d\theta}{\theta^3}$ divergence that appears
in the cross section becomes softer as schematically 
$dE/dx \sim n \sigma E^\prime$, where, $E^\prime/E=\frac{1}{2}(1-\cos\theta) 
\sim \theta^2/4$, tames the divergence. This cut-off procedure 
can be avoided by incorporating appropriate screening effects
\cite{thoma91,braaten91,romatschke}. The differential collision probability
is defined as 
\bea
\p=\frac{\pi\alpha_s^2}{2E^2}\int\frac{d^3k}{k}\mid {\cal{M}}\mid^2 f(k)
\label{theta}
\eea
where
\be 
f=\frac{\nu}{(2\pi)^3}\frac{1}{exp(k/T)\pm 1}
\ee 
where $\nu$ is the statistical degeneracy of the particles in 
the thermal bath.
%Here $t=-sE^\prime/E$,
%$u=-s(1-E^\prime/E)$ which ensures $s+u+t=0$.
It might be mentioned that for coulomb like scattering, $\th(E,E^\prime)
\propto 1/E^{\prime 2}$ and therefore
$dE/dx \propto \int \frac{dE^\prime}{E^\prime}$. This, evidently, is divergent
and gives the logarithmic dependence of $dE/dx$ \cite{bjorken}. We also 
assume that the energy of the incoming parton $E>>T$, where $T$ is the
temperature of the system.

Now we consider a specific process $ q q^\prime \rightarrow q q^\prime$,
the matrix element for which is given by the following expression,

\bea
|{\cal{M}}|^2= \frac{4}{9} g^4 \frac{s^2+u^2}{t^2}
\label{matqq}
\eea
where $g^2=4\pi\alpha_s$ is the color charge and $\alpha_s$ 
is the strong coupling constant.

For $ q q^\prime \rightarrow q q^\prime$,
the differential collision probability can be expressed as
\bea
\th(E,E^\prime)=\frac{\pi\alpha_s^2}{18}\frac{T^2}{E^2}(1-2\frac{E}{E^\prime}+
2\frac{E^2}{E^{\prime 2}}) 
\eea
where $T$ is the temperature of the medium.

The last equation together with Eq.~\ref{dedx} gives the following
expression for the energy loss,

\bea
\frac{dE}{dx}\approx\frac{\pi\alpha_s^2 T^2}{9} ln(E/\wpl_0)
\label{dedxhard0}
\eea
where $\wpl_0$ is the lower cut off. In ref.~\cite{bjorken} this is
taken as $\mu^2/2k$ while in ref~\cite{baier00} $\wpl_0\,\sim\,\alpha_sT$.
We in section IV, show that how this cut off follows from the full
calculation.  A complete treatment of this singularity
in the context of heavy quark energy loss both for soft and hard momentum
transfer within hard thermal loop re-summation scheme has been discussed
in \cite{braaten91}. We present corresponding calculations for the light 
quarks in a slightly different approach.
  
%To deal
%with the hard momentum exchange we take the approach of Ref.~\cite{bjorken} 
%where to regulate the divergence an infrared cutoff $(\omega_0)$ is used. 
%In addition, we add the contributions coming from the soft momentum
%transfer for which hard thermal loop~\cite{pisarski} 
%resummed propagators are used. This, in effect, removes the arbitrariness
%of the lower cut-off (see section IV for details).
%
In the subsequent sections, we present explicit expressions of $\p$ for
various processes. Corresponding QED results are also derived for
comparison \cite{rossi1,rossi2}. The quantities $E$ and $E^\prime$ are defined
in the laboratory frame.

\section{Contribution of various channels} 

\subsection{Quarks} 

Let us first consider propagation of a hard quark through 
a QCD plasma. The collisions which would contribute to its energy loss are
$q q \rightarrow q q$, $ q q^\prime \rightarrow q q^\prime $, 
$ q \bar{q}^\prime \rightarrow q \bar{q}^\prime $, 
$q g \rightarrow q g$ and $q \bar{q} \rightarrow q \bar{q},gg, 
q^\prime \bar{q}^\prime$. In the above processes primes indicate 
different flavors.  It might be mentioned that 
in a baryon free region, {\em i.e.}, in absence of a net baryonic
chemical potential, quark and anti-quark energy loss will be the
same. Therefore, we do not treat them separately.

The most dominant process for quark energy loss, as mentioned before, 
is compton like scattering, {\em i.e.} $ q g \rightarrow q g$. Beside the
$t$ channel, contributions from additional diagrams are found to be
non-negligible. 

The differential collision probability for the compton like scattering
is as follows:
 
\bea
\th_{q g \rightarrow q g}(E,E^\prime)dE^\prime
&=&  \frac{2\pi \alpha_s^2}{3 E^2}\,T^2\big [
1- 2  \frac{E}{E^\prime} + 2 (\frac{E}{E^\prime})^2\nn\\
&+& \frac{4}{9}
\{ 1 - \frac{E^\prime}{E} + \frac{E}{E-E^\prime}
\} 
\big ]dE^\prime
\label{compton}
\eea

In Eq.~\ref{compton}, the first three terms come from the $t$ channel and
others originate from the exchange diagram and $s$ channel. Evidently,
$E/(E-E^\prime)$ gives rise to logarithmic enhancement. Thus overall
contribution becomes significant if one retains all the possible
diagrams. To get the energy loss, one integrates the differential
collision probability weighted with the energy transfer as shown in
Eq. \ref{dedx} to yield :  
\be
(\frac{dE}{dx})_{q g \rightarrow q g } 
=\frac{2\pi \alpha_s^2}{3} \,T^2
\big [\frac{22}{9} \ln(E/\wpl_0) -0.176
\big]
\label{comptondedx}
\ee
In writing the above equation, we have used $E_{\mathrm{min}}=\wpl_0/2$,
and $E_{\mathrm{max}}=E$, the energy of the hard parton, it is also
assumed here that $E>>\wpl_0$.
This is justified, as for the present problem, 
partonic jets have very high energy compared 
to the energy of the plasma constituents which could be $\sim$ 3T.
It should be noted that the coefficient of the logarithmic term
is different from that one obtains by restricting
to the $t$ channel alone.  

Next we consider M\"oller type $q q \rightarrow q q $ scattering for which
the differential collision probability \cite{alam} reads,

\bea
\th_{qq\rightarrow qq}(E,E^\prime)dE^\prime
&=&
\frac{\pi \alpha_s^2}{9} \,T^2
\big [ \frac{E^2}{E^{\prime 2}(E-E^\prime)^2}\nn\\ 
&+& 
\frac{\Delta}{E^\prime (E -E^\prime) } + \frac{1}{E^2} \big ]dE^\prime,
\eea
where, $\Delta = -10/3$. 
In the last expression if we replace $\frac{2}{9}\alpha_s^2$ by
$\alpha_{em}^2$ and $\Delta$ by -2, the electron energy loss due 
to M\"oller scattering ensues \cite{rossi1,rossi2}. 
This reaction deserves special attention as it involves two identical
particles in the final state. 
Therefore, $\p$, in this case, should be 
interpreted as the probability of a collision which leaves one parton 
in the energy state $E^\prime$ and the other in the energy state
$E-E^\prime$. To take into account all the possibilities, $E^\prime$ is
varied from $\wpl_0/2$ to $E/2$ \cite{rossi1,rossi2}. Similar subtlety is involved
for processes like $gg \rightarrow gg$ or $q \bar{q} \rightarrow gg$ etc. 
The final expression for $q q \rightarrow q q $ is given by 
\be
\left ( \frac{dE}{dx} \right )_{qq \rightarrow q q}
=\frac{\pi\alpha_s^2}{9} \,T^2
\big [
\ln(E/\wpl_0)+\Delta\ln 2 +1.125 
\big ].
\label{mollerdedx}
\ee

The other important reaction is $ q \bar{q} \rightarrow q \bar{q}$,
which also has  $t^{-2}$ divergence~\cite{leader} and therefore, found to 
contribute significantly to the total energy loss. It should be
noted that there is no $u^{-2}$ divergence involved in this process
hence the collision is dominated by soft scattering and result
do not differ much if the relevant $s$ channel diagram is excluded.
We, nevertheless, retain all the diagrams. The differential probability
for `bhabha' like scattering, therefore, takes the form :
\bea
\th_{q \bar{q}\rightarrow q \bar{q}}(E,E^\prime)dE^\prime
&=&
\frac{\pi\alpha_s^2}{9 {E^\prime}^2}\,T^2
\big [1-\Delta^\prime\frac{E^\prime}{E}+(2\Delta^\prime-1)
\frac{{E^\prime}^2}{E^2}\nn\\
&&-
\Delta^\prime\,\frac{{E^\prime}^3}{E^3}+
\frac{{E^\prime}^4}{E^4}
\big]dE^\prime
\eea
Corresponding energy loss turns out to be :
\be
(\frac{dE}{dx})_{q \bar{q} \rightarrow q \bar{q}}
=
\frac{\pi \alpha_s^2}{9} \,T^2
\big [\ln{\frac{E}{\wpl_0}}-\frac{\Delta^\prime}{3}+0.443
\big ],
\ee
where $\Delta^\prime = 2/3$. The QED limit for the last two 
equations can be taken by replacing $\frac{2}{9}\alpha_s^2$ 
with $\alpha_{em}^2$ and $\Delta^\prime = 2$~\cite{rossi1,rossi2}.

Finally we present results for the process, $q\bar{q}\rightarrow gg$.
This again involved identical particles in the
final channel for which appropriate limit is taken. This process
is also suppressed because of less sensitive infrared divergences 
as evident from the expression:

\bea
\th_{q \bar{q}\rightarrow g g}(E,E^\prime)dE^\prime
&=&\frac{\pi \alpha_s^2}{3 E^2} \,T^2  
\big [
\frac{4}{9} \{ \frac{E}{E^\prime} + \frac{E^\prime}{E-E^\prime} \}
+ 2 \frac{E^\prime}{E}\nn\\ 
&-&2 \frac{E^{\prime 2} }{E^2} 
-\frac{13}{9} 
\big ]dE^\prime
\label{qqbargg}
\eea

Other important reactions for which we do not present explicit
results include $q q^\prime\rightarrow q q^\prime$ and 
$q \bar{q}^\prime\rightarrow q \bar{q}^\prime$. They
contribute equally to the energy loss (for baryon free matter).
It should be mentioned that $q \bar{q} \rightarrow q^\prime \bar{q}^\prime$
induced energy loss is small because of the absence
of infrared enhancement.  This is less
divergent (no $t^{-2}$ or $u ^{-2}$), and, therefore, 
found to be less effective
means of energy dissipation. 

\subsection{Gluons} 

Similar to quarks, hard gluons can also dissipate energy while colliding
with the plasma constituents. Most important process by which gluons can
transfer energy to the plasma is the $g g \rightarrow g g$. 
\bea
\th_{g g\rightarrow g g}(E,E^\prime)dE^\prime
&=&\frac{3\pi \alpha_s^2}{E^2} \,T^2  
\big [
3 - \frac{E^\prime(E-E^\prime)}{E^2} 
+  \frac{E^2}{E^{\prime 2}}\nn\\ 
&-& \frac{E}{E^\prime}  
+ \frac{E E^\prime}{(E-E^\prime)^2} 
\big ]dE^\prime
\eea
\begin{figure}%[*htb]
\epsfig{figure=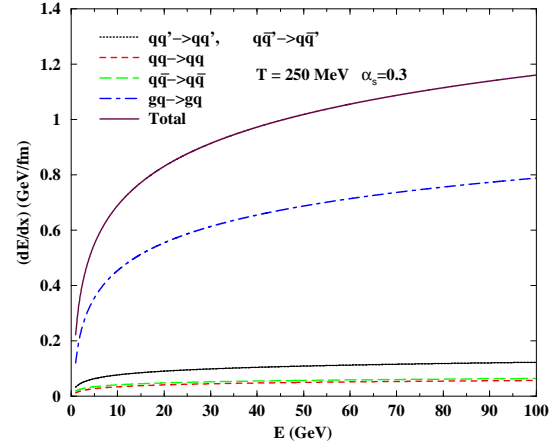, height= 6 cm}
\caption{ Individual contributions of various processes responsible for
quark energy loss are shown. The aggregated collisional loss is also presented} 
\label{fig1}
\end{figure}

Corresponding expression for the energy loss can be written as 
\be
(\frac{dE}{dx})_{gg\rightarrow gg } 
=3\pi \alpha_s^2\,T^2 
\big[\ln(\frac{E}{\wpl_0}) -0.038
\big]
\ee
Like quark, the QCD compton scattering also proves to be quite efficient in
transferring gluon energy into the plasma. Relevant expressions for the
differential collisional probability $\p$ and 
$dE/dx$ induced by $g q \rightarrow g q$ scattering can be obtained from
Eqs. \ref{compton} and \ref{comptondedx} respectively by appropriately
replacing the phase space factor
(factor 2/3 in eqs.~\ref{compton} and \ref{comptondedx} should be replaced
by  3/4 for three flavour QGP).
It should be mentioned that $ g g \rightarrow q \bar{q}$ is
also suppressed as there is no $t^{-2}$ or $u^{-2}$ singularity involved
in this process. Gluonic energy loss induced by this process can be
obtained from Eq.~\ref{qqbargg}.

In Fig.~\ref{fig1} we present stopping power as 
function of energy of the incoming 
parton at a temperature $T = 250$ MeV. The result is to be compared  
with previous estimates \cite{bjorken,baier00}. 
Evidently, bulk contribution to the total
collisional energy loss of quark comes from the 
$ q g \rightarrow q g $ channel.  Net energy loss of a light
quark is given by the sum of all these diagrams including scattering
and annihilation processes.  
Contribution of inelastic channels 
are found to be small and, therefore, have not been shown explicitly. 
However, the total loss, as demonstrated in Fig.~\ref{fig1} include effect of
all the channels.  It
should be mentioned that present treatment can  be extended
for heavy quarks for which collision probabilities will be 
modified \cite{alam}. Quantitatively, we find  $dE/dx \sim 0.8$ GeV/fm 
for a $20$ GeV parton, {\em vis-a-vis} $0.2$ GeV/fm of 
Refs.\cite{bjorken,baier00,braaten91}. This can be attributed to the
diagrams other than $t$ channel. 

The results for gluon energy loss is presented in Fig.~\ref{fig2} below. 
Evidently
gluon energy loss is mostly driven by $g g \rightarrow g g$ scattering. 
Also comparable is the contribution of $ q g \rightarrow q g$ channel.

\begin{figure}
\epsfig{figure=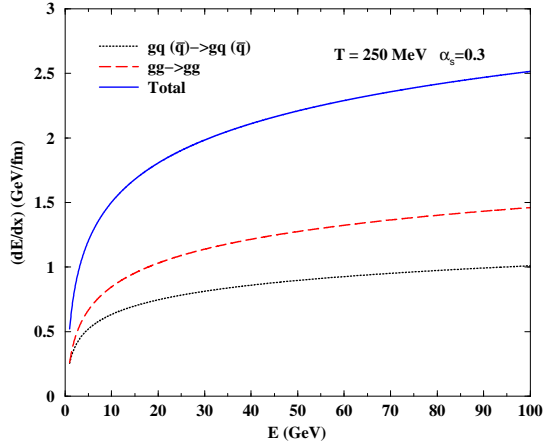, height= 6 cm}
\caption{ Individual contributions of various processes responsible for
gluon energy loss are shown. Dashed, dot-dashed and solid line represent
$g q({\bar q}) \rightarrow g q({\bar q})$,
$g g \rightarrow g g$
and total respectively.}
\label{fig2}
\end{figure}

\subsection{Comparison with radiative loss}

\begin{figure}%[*htb]
\epsfig{figure=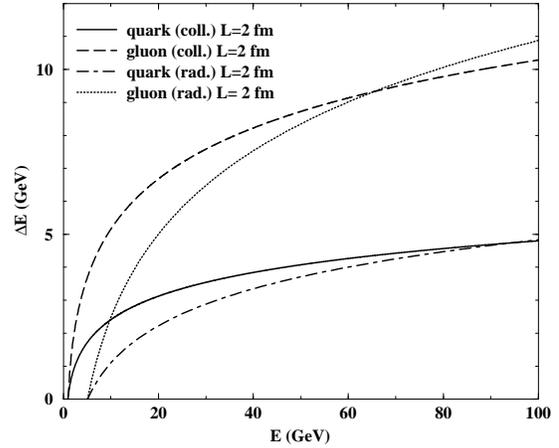,height=6 cm}
\caption{The collisional  and radiative 
energy loss of quarks and gluons passing through quark gluon plasma.
}
\label{fig3}
\end{figure}

To bring the importance of collisional loss into bold relief, we 
estimate the possible parton density relevant for the RHIC energies.
The gluon rapidity density in this case can be taken as 
$dN_g/dy \sim 1000$, which, when plugged into the Bjorken formula~\cite{jdb83}
$\rho_g = \frac{dN_g}{dy}/\tau_0\pi R_{Au}^2$ with formation time 
$\tau_0=0.5$ fm/c, we get a value of $T \sim 400$ MeV. 
It might be mentioned that
this density is consistent with the one used in Ref.\cite{gyulassy00prl}.
Corresponding values of the total (integrated over plasma length) energy loss
for quark and gluon is significantly large as depicted in Fig.~\ref{fig3}. 
We also compare total collisional energy loss with its radiative
counterpart.  
%Similar to QED~\cite{leo}, we determine the critical energy 
%$E_c$ from the condition: $(dE/dx)_{rad}=(dE/dx)_{coll}$.
%Evidently for parton energy $< E_c\sim 100$ GeV
%the collisional loss seems to be more important and for higher energies
%the bremsstrahlung radiation becomes dominant. 
For the latter, we take~\cite{radloss}
\be
\Delta E_{rad}= C_2\frac{\alpha_s\mu^2 L^2}{N(E)\lambda } 
\ln(\frac{2E}{\mu^2L})
\ee
where $L$ is the length of the plasma traversed by the partons and
$\lambda$ is the mean free path.
Collisional energy loss is more than its radiative counterpart for
parton energy up to $E=E_c \sim 85 $ GeV for quarks and $60$ GeV for
gluons repetitively. The results shown in Fig~\ref{fig3} correspond to
 $N(E)=10 $, $\mu=1$ GeV, $\alpha_s=0.3$  and $L/\lambda=4$.
It is important to point out here that 
$N(E)=7.3, 10.1, 24.4$ for $E=500, 50, 5 $ GeV 
respectively and $N(E\rightarrow\infty)=4$~\cite{gyulassy00prl}.

\begin{figure}%[*htb]
\epsfig{figure=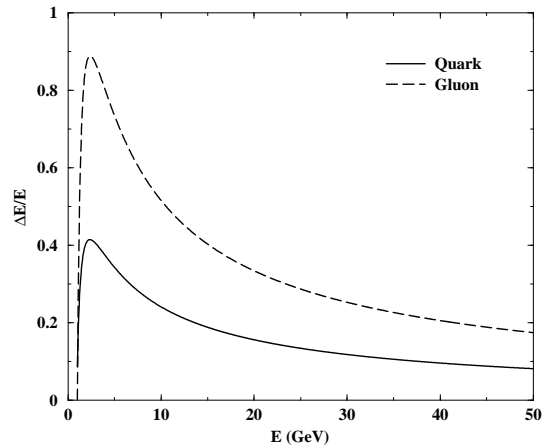,height=6 cm}
\caption{The fractional energy loss of quarks and gluons due to collisions.
The length of the medium $L=2$ fm, $T=400$ MeV.
}
\label{fig4}
\end{figure}

In Fig.~\ref{fig4} we depict the variation of fractional energy loss
due to the collision of quarks and gluons passing through a QGP
medium of length $L=2$ fm at $T= 400$ MeV. This result may be contrasted
with its radiative counter part as given in~\cite{lpf02}. It is important
to point out here that a value of $\Delta E/E \sim 1/5$ ~\cite{shuryak} 
can reproduce the high $p_T$ suppression of pion spectra observed at RHIC energy.

\begin{figure}%[*htb]
\epsfig{figure=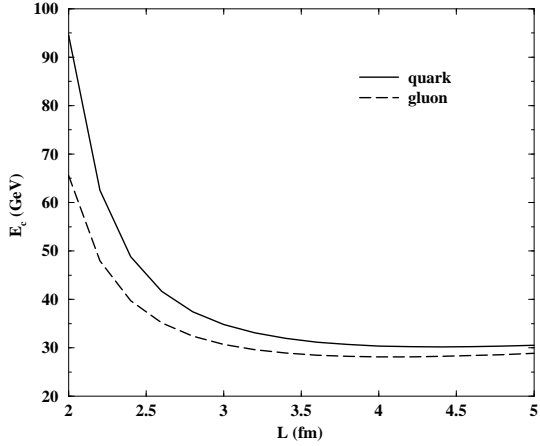,height=6 cm}
\caption{The variation of $E_c$ as function of path length 
of the high energy partons moving through  the QGP at 
$T=400$ MeV.
}
\label{fig5}
\end{figure}

In Fig.~\ref{fig5} we show the variation of the critical energy
of the high energy parton as a function of its path length
in the medium where the radiative and the collisional
losses contribute equally. 
Here the results  are obtained for 
$N(E)=10 $, $\mu=1$ GeV, $\alpha_s=0.3$  and $\lambda=0.5$ fm.

\section{Cancellation of the intermediate scale}

In the previous section the results for the quark and gluon energy loss
are presented with a lower cut off $\wpl_0$.  This was introduced to regulate
the infrared divergence. In this section, we explicitly show that infrared
divergence is automatically screened by the plasma effects while there exists
a separation scale coming from the two kinematic regimes. Ultimately it
is removed from the
final expression once  both the hard scattering and the collective
plasma effects are added. We also assume that the energy of the
incoming parton is much larger than the temperature, 
{\it i.e.} $E>>T$. The soft
and hard momentum transfer ($q$) is determined by the fact whether $q \sim gT$
or $q \sim T$ and while for the former the interaction is screened by
the Debye mass (for the electric mode or dynamically for the magnetic
part). The intermediate scale $q^*$ is chosen such that $gT<q^*<T$. We,
then show that the final result is independent of this scale $q^*$. 

Following Braaten {\it et al.}~\cite{braaten91}, 
we define the energy loss $dE/dx$ 
as a product  of collision rate and energy transfer per scattering divided by
the velocity \cite{braaten91} :
\bea
\frac{dE}{dx}=\frac{1}{v}\int dE_3(E-E_3)\frac{d\Gamma}{dE_3}
\eea
where $\Gamma$ is the interaction rate.
The equivalence with the previously defined $\th(E,E^\prime)$ can be 
established easily by identifying $\th(E,E^\prime)=v^{-1}d\Gamma/dE^\prime$, 
where $E^\prime=E-E_3$ is the energy transfer.

\bea 
\frac{dE}{dx}&=&\frac{\nu}{2E}\int 
\frac{d^3p_1}{2E_1(2\pi)^3}
\frac{d^3p_2}{2E_2(2\pi)^3}
\frac{d^3p_3}{2E_3(2\pi)^3}\frac{(E-E_3)}{v}\nn\\
&&\left [ f_1 (1-f_2)(1-f_3)\pm (1-f_1) f_2 f_3 \right ]\nn\\
&&(2\pi)^4 \delta^4(P+P_1-P_2-P_3) |{\cal{M}}|^2
\label{dedxdef}
\eea
Note the difference of the thermal phase space here with that of the 
heavy quarks where only the factor $f_1(1-f_2)$ appears~\cite{braaten91}
due to the absence of heavy quarks in the thermal bath, deleting
the possibility of reverse reactions.
In the above equation, $\nu$ stands for the statistical degeneracy
factor. For the processes under consideration $q q^\prime \rightarrow 
q q^\prime$, we have the following matrix element 

\bea
|{\cal{M}}|^2&=&4 g^4C_{qq} D_{\mu\nu}(q) D_{\alpha\beta}^\ast(q) 
%\left 
[ (P^\mu P_3^\alpha + P_3^\mu P^\alpha - g^{\mu\alpha}(P.P_3))\nn\\
&&(P_1^\nu P_2^\beta + P_2^\nu P_1^\beta - g^{\nu\beta}(P_1.P_2))
\label{matelfull}
%\right ]
\eea
where $C_{qq}$ is the colour factor.
The matrix element Eq.~\ref{matelfull}, in general is very complicated,
which takes a simple form in the limit of soft momentum transfer or
small angle scatterings. This is justified because  of the
infrared sensitivity, energy loss is dominated by the soft collisions.

In the coulomb gauge, we can define $D_{00}=\Delta_l$ and 
$D_{ij}=(\delta_{ij}-q^i q^j/q^2)\Delta_t$.
$\Delta_l$ and $\Delta_t$ denote the longitudinal and transverse 
gluon propagators given by,
\bea
\Delta_l(q_0,q)^{-1}=q^2-\frac{3}{2}\omega_p^2 
\left [
\frac{q_0}{q}ln\frac{q_0+q}{q_0-q}-2
\right ]
\eea

\bea
\Delta_t(q_0,q)^{-1}=q_0^2-q^2+\frac{3}{2}\omega_p^2 
\left [\frac{q_0(q_0^2-q^2)}{2q^3}
ln\frac{q_0+q}{q_0-q}-\frac{q_0^2}{q^2}
\right ]
\eea

With this, matrix element in the limit of small angle scattering,
for which $P.P_1=P_2.P_3 >> P.P_3$ or $P_1.P_4$, we get the following
expression for the squared matrix element.

\bea
|{\cal{M}}|^2&=& g^4 C_{qq} 16 (E E_1)^2 
\vert \Delta_l(q_0,q)  \nn\\
&+& (v\times {\hat{q}}).(v_1 \times \hat{q}) 
\Delta_t(q_0,q)\vert^2
\eea

with $v=\hat{p}$ and $v_1=\hat{p_1}$. We also use energy conservation
\bea
q_0=E-E_3=E_2-E_1 \nn
\eea
this in the soft limit, {i.e. $q<<T$}, becomes
\bea
q_0\simeq v\cdot q \simeq v_1\cdot q
\eea
Another useful identity that helps to cast Eq.\ref{dedxdef} in a 
simplified form is the following
\bea
&&\int 
\frac{d^3p_2}{(2\pi)^3}
\frac{d^3p_3}{(2\pi)^3}
(2\pi)^4 \delta^4(P+P_1-P_2-P_3)\nn\\
&&\simeq
\int \frac{dq_0d^3q}{2\pi (2\pi)^3}
2\pi \delta(q_0-v.q) 2\pi \delta(q_0-v_1.q)
\eea

These delta functions can be used to perform the angular integrations,
while the integration over $p_1$ can be obtained by means of partial
integration,

\bea
&&\int dp_1 p_1^2 \left (-\frac{df_1}{dp_1} \right )\nn\\
&&=2 \int dp_1 p_1 f_1= \frac{\pi^2T^2}{6} 
\eea

In case of fermionic initial and final states
we have
\bea
&&f_1 (1-f_2)(1-f_3) + (1-f_1) f_2 f_3\nn\\  
&&=(f_1-f_2)\left [ 1 + N(q_0) - f_3 \right ]\\
&&\simeq -\frac{df_1}{dp_1}q_0 \left[\frac{T}{q_0}-\frac{1}{2} \right ]
\eea

Here $N(q_0)=(exp(q_0/T)-1)^{-1}$.

\bea
&&\frac{dE}{dx}=\frac{\nu g^4}{2E}\int 
\frac{d^3p_1}{2E_1(2\pi)^3} 
\frac{d^3q}{(2\pi)^3} 
\frac{dq_0}{2\pi} 
\frac{df_1}{dp_1} q_0^2\left (-\frac{T}{q_0}+\frac{1}{2}
 \right )\nn\\
&&\frac{1}{4 E_2 E_3}|{\cal{M}}|^2 2\pi \delta(q_0-v.q) 2\pi \delta(q_0-v_1.q)
\eea
It might be noted that for the calculation of collisional
rate, $\Gamma$, the term $T/q_0$ contributes at the
leading order, while for the $dE/dx$ the with 1/2
inside the bracket gives non-zero contribution. This is
related to the parity of the spectral function of the
gluons. 

Final expression for the energy loss is given by
\bea
\frac{dE}{dx}&=&\frac{g^4C_{qq}
T^2}{96 \pi}\nu \int dq \int_{-q}^q q_0^2 dq_0\nn\\
&&\left[ |\Delta_l|^2 + \frac{1}{2}(1-\frac{q_0^2}{q^2})^2|\Delta_t|^2
\right ]
\label{elosseq1}
\eea

Before proceeding further, let us see what happens if we use the bare
propagator in the last equation which is given by 
$|\Delta_l(q_0,q)|^2=1/q^4$ and $|\Delta_t(q_0,q)|^2=1/(q_0^2-q^2)^2$. With
this the expression for $dE/dx$ turns out to be

\bea
-\frac{dE}{dx}
&=&\frac{g^4T^2}{96\pi}\nu C_{qq}\int_{q^*}^{T} \frac{dq}{q}\nn\\
&=&\frac{g^4T^2}{96\pi}\nu C_{qq}ln\frac{T}{q^\ast}
\label{dedxhard}
\eea 
which clearly is logarithmically  divergent. The upper limit indicates
break down of the approximation beyond $T$ and the lower limit
is to regulate the infrared divergence. 

It is instructive to compare this with the corresponding limit of the 
collision rate which diverges quadratically~\cite{blaizot02}
\bea
\Gamma=\frac{g^4T^2}{96\pi}\nu C_{qq}\int \frac{dq}{q^3}
\eea 

Eq.\ref{elosseq1}, can be expressed in terms
of the spectral functions and directly be compared with ref.~\cite{braaten91}.
For this we recall that the transverse and longitudinal propagators have the
following spectral representations,

\bea
\Delta_l(q_0,q)=-\frac{1}{q^2} + \int_{-\infty}^\infty 
\frac{d\omega}{2\pi}\frac{\rho_l(\omega,q)}{\omega-q_0}
\eea

\bea
\Delta_t(q_0,q)=\int_{-\infty}^\infty 
\frac{d\omega}{2\pi}\frac{\rho_t(\omega,q)}{\omega-q_0}
\eea
where,
\bea
\rho_{l,t}=2 Im \Delta_{l,t}(q_0+i\epsilon,q)
\eea
The spectral function contains contributions both from the residue
at the pole and the discontinuity due to the branch cuts,
\bea
\rho_{l,t}(q_0,q)&=&2\pi \epsilon(q_0)z_{l,t}(q) \delta[q_0^2-\omega^2_{l,t}(q)]\nn\\
&+& \beta_{l,t}(q_0,q)\theta(q^2-q_0^2). 
\eea
Here $z_{l,t}(q)$ is the residue
of the time like pole at $\omega_{l,t}$ and $\beta_{l,t}$ is the
contribution from the branch cuts.

\bea
\beta_{l}(q_0,q)=3\pi\omega_p^2 \frac{q_0}{q} |\Delta_l(q_0,q)|^2,
\eea

\bea
\beta_{t}(q_0,q)=3\pi\omega_p^2 \frac{q_0(q^2-q_0^2)}{2q^3} |\Delta_t(q_0,q)|^2
\eea

With these equations, the energy loss can be expressed as
\be
-\frac{dE}{dx}=\frac{g^2C_{qq}\nu}{16\pi}
\int qdq \int_{-q}^{+q}\frac{q_0dq_0}{2\pi}
\left[\rho_l(q_0,q)\nn\\
+(1-\frac{q_0^2}{q^2})\rho_t
\right ]
\ee

To calculate the soft part, we make use of the identities listed in
the appendix for $\omega_p < q^* < T$.

\bea
-\frac{dE}{dx}&=&
\frac{g^2C_{qq}\nu}{16\pi}
\int_{\omega_p}^{q^\ast}q dq
\left [ I^l_{(1)} + I^t_{(1)} - \frac{1}{q^2}I^t_{(3)}
\right ]\nn\\
&=& \frac{3\nu g^2C_{qq}}{32\pi}\omega_p^2ln\frac{q^*}{\omega_p}
\label{dedxsoft}
\eea

It might be mentioned that the hard part can also be calculated from 
Eq.~\ref{dedxsoft} by taking
\bea
\rho_l(q_0,q) \simeq \frac{3\omega_p^2 q_0}{2 q^5}\\
\rho_t(q_0,q) \simeq \frac{3\omega_p^2 q_0}{4 q^5(1-\frac{q_0^2}{q^2})}\\
\eea
yielding,
\bea
-\frac{dE}{dx}=\frac{3\nu g^2 C_{qq}}{32\pi}\omega_p^2ln\frac{T}{q^\ast}
\eea
 Eq.~\ref{dedxsoft} and the last equation clearly shows that the intermediate
scale gets canceled when both the contributions are added together. 

In the present treatment, to show the cancellation, we focussed only on
the leading log part. However, full calculation can be done along the line 
of~\cite{braaten91} with appropriate 
modification of the kinematics, {\em i.e.} $q_0 \leq p - p_1$ and
$q \leq q_0 + 2 p_1$. Using the fact that $p=E>>T$, one finds 
$q_{max} \simeq E$. With these, by adding the soft and hard 
contributions, one obtains
\bea
-\frac{dE}{dx}=\frac{3\nu g^2 C_{qq}}{64\pi}\omega_p^2ln\frac{E}{g^2T}
\eea 
This expression coincides with Eq.~\ref{dedxhard0}  with 
appropriate degeneracy and colour factors. 
 
\section{Summary and Discussions}
To summarize, in the present work, we have studied 
collisional loss of light partons in  hot QCD plasma.
We have identified  some of the important
diagrams  previously ignored which include $u$ channel 
divergences contributing to the leading log results.
Subtleties related to the identical
final state particles which were overlooked earlier 
(leads to overestimation in $dE/dx$)  
in dealing with light quarks and gluons 
This has properly been taken into account in the present work.
Our results are free from any arbritrary cut off that was
present in \cite{bjorken,mrow91,koike}.
Moreover, the conditions where the collisional and the radiative losses
are comparable is clearly revealed. Furthermore, note that RHIC data 
suggests only a tiny amount of 
`quenching', $Q_t(p_T)$=0.2\cite{shuryak}. This  corresponds to a 
small amount of
energy loss which might be accommodated with  the collisional
loss. In addition, collisional energy loss has a 
different qualitative importance as it injects energy into the plasma.
Implication of this has recently been discussed in Ref~\cite{shuryak}. 

One of the authors (AKDM)  would like to thank S. Mrowczynski and I. Vitev for
useful discussion.

%\addcontentsline{toc}{section}{Appendix : I}
\section{Appendix}
\setcounter{equation}{0}
\def\theeuqation{I.\arabic{equation}}
In this appendix we present various sum rules used in this work 
(see ~\cite{bellac} for more details).
Expressions for the spectral sum rules :\\
\bea
&&\int_{-q}^{q} \frac{dq_0}{2\pi} \frac{\beta_l(q_0,q)}{q_0}
=\frac{1}{q^2}-\frac{1}{q^2+3\omega_p^2} - \frac{z_l(q)}{\omega_l^2(q)}\\
&&\int_{-q}^{q} \frac{dq_0}{2\pi} q_0\beta_l(q_0,q)
=\frac{\omega_p^2}{q^2}-z_l(q)\\
&&\int_{-q}^{q} \frac{dq_0}{2\pi} q_0^3\beta_l(q_0,q)
=\frac{3}{5}\wpl_p^2+ \frac{\omega_p^4}{q^2}-z_l(q)\omega_l^2(q)
\eea

where,
\bea
z_l(q)=\frac{2\omega_l^2(\omega_l^2-q^2)}{q^2(3\omega_p^2+q^2-\omega_l^2)}
\eea

\bea
&&\int_{-q}^{q} \frac{dq_0}{2\pi} \frac{\beta_t(q_0,q)}{q_0}
=\frac{1}{q^2} - \frac{z_t(q)}{\omega_t^2(q)}\\
&&\int_{-q}^{q} \frac{dq_0}{2\pi} q_0\beta_t(q_0,q)
=1-z_t(q)\\
&&\int_{-q}^{q} \frac{dq_0}{2\pi} q_0^3\beta_t(q_0,q)
=q^2 + \omega_p^2 -z_t(q)\omega_t^2(q)
\eea

where,
\bea
z_t(q)=\frac{2\omega_t^2(\omega_t^2-q^2)}{3\omega_p^2\omega_t^2-
(\omega_t^2-q^2)^2}
\eea

\bea
z_l(q)_{q\rightarrow 0}&=&\frac{\omega_p^2}{q^2},\\
z_l(q)_{q\rightarrow \infty} &=&0\\
\eea

\bea
z_t(q)_{q\rightarrow 0}&=& 1 -\frac{q^2}{5\omega_p^2}, \\
z_t(q)_{q\rightarrow \infty } &=&1+\frac{3\omega_p^2}{4q^2}\\
\eea

Similarly, 
\bea
\omega_l^2(q)_{  q\rightarrow 0} &=& \omega_p^2+\frac{3}{5}q^2\\
\omega_l^2(q)_{ q\rightarrow \infty }&=& q^2 [1+4e^{-
\frac{2}{3}\frac{q^2}{\omega_p^2} -2}]
\eea
\bea
\omega_t^2(q)_{ q\rightarrow 0} &=& \omega_p^2+\frac{6}{5}q^2\\
\omega_t^2(q)_{ q\rightarrow \infty }&=& q^2 + \frac{3}{2}\omega_p^2
\eea

With the help of these limiting values one can readily evaluate 
following integrals which can be used to evaluate the integrals
involving $\rho_{l,t}$ as appear in the expression for the energy
loss.

Defining 
\bea
I^{l,t}_{(n)}=\int_{-q}^{+q} 
\frac{dq_0}{2\pi}\beta_{l,t}(q_0,q)q_0^n
\eea

For $q>>\omega_p$

Longitudinal
\bea
I^l_{(-1)}&=& \frac{3\omega_p^2}{q^4}\\
I^l_{(1)}&=& \frac{\omega_p^2}{q^2}\\
I^l_{(3)}&=&\frac{3}{5}\omega_p^2+\frac{\omega_p^4}{q^2}\\
\eea
Transverse
\bea
I^t_{{(-1)}}&=&\frac{3\omega_ p^2}{4q^4}\\
I^t_{(1)}&=&-\frac{3\omega_p^2}{4q^2}\\
I^t_{(3)}&=&-\frac{5}{4}\omega_p^2
\eea

For $q<<\omega_p$
Longitudinal
\bea
I^l_{(-1)}&=& \frac{4}{15}\frac{1}{\omega_p^2}\\
I^l_{(1)}&=& 0\\
I^l_{(3)}&=& 0\\
\eea
Transverse
\bea
I^t_{(-1)}&=& \frac{1}{q^2}-\frac{1}{\omega_p^2}\\
I^t_{(1)}&=& \frac{q^2}{5\omega_p^2}\\
I^t_{(3)}&=& 0\\
\eea

\end{multicols}
\end{document}